\documentclass{article}

\usepackage{arxiv}

\usepackage[utf8]{inputenc} 
\usepackage[T1]{fontenc}    
\usepackage{hyperref}       
\usepackage{url}            
\usepackage{booktabs}       
\usepackage{amsfonts}       
\usepackage{nicefrac}       
\usepackage{microtype}      
\usepackage{lipsum}		
\usepackage{graphicx}
\usepackage{epstopdf, epsfig}
\usepackage{amsmath}
\usepackage{newtxtext}
\usepackage{newtxmath}
\usepackage{natbib}
\usepackage{hyperref}
\usepackage{siunitx}
\usepackage{float}
\usepackage[subrefformat=parens]{subcaption}
\usepackage{natbib}
\usepackage{doi}

\title{A fast immersed boundary method for an extruded wall geometry}

\author{ {Manabu Saito} \\
	Department of Mechanical Engineering and Science\\
	Kyoto University\\
	Kyoto daigaku-Katsura, Nishikyo-ku, Kyoto 615-8540, Japan \\
	\texttt{saito.manabu.63z@st.kyoto-u.ac.jp} \\
	\And
	{Ryoichi Kurose} \\
	Department of Mechanical Engineering and Science\\
	Kyoto University\\
	Kyoto daigaku-Katsura, Nishikyo-ku, Kyoto 615-8540, Japan \\
	\texttt{kurose@mech.kyoto-u.ac.jp} \\
}

\date{}

\renewcommand{\shorttitle}{A fast IBM for an extruded wall geometry}

\begin{document}
\maketitle

\begin{abstract}
The computational cost of the boundary-condition-enforced immersed boundary method (IBM) increases in the order of $\mathcal{O}(N^2)$ as the number of Lagrangian points, $N$, increases.
This is due to the time-consuming calculation of the correction operator in the diffuse-interface IBM to enforce the no-slip boundary condition.
In this study, a computationally efficient IBM algorithm for an extruded wall geometry is developed, and the correction operator calculation is significantly simplified while maintaining the accuracy of the solution.
This method takes advantage of the geometrical symmetricity to apply several matrix simplifications, which result in a huge increase in the computational efficiency and an improved scalability of $\mathcal{O}(max(N, N^2/r^2))$ ($r$: the number of grid points towards the extruded direction).
The boundary-condition-enforced IBM for an extruded wall geometry is applicable to the numerical simulations of the flow around a wall surface that satisfies both (a) an extruded wall geometry that retains the same cross-section geometry and (b) an Eulerian grid that is either uniform or whose stretch rate towards the extruded direction is constant.
As this type of geometry is commonly studied to investigate the fundamental behavior of the fluid, the presented algorithm has wide applications.
Several calculations are conducted to demonstrate the higher computational efficiency of the presented algorithm compared with that of the original algorithm.
The results show improvements in computational efficiency of up to 2,800 times for the correction operator calculation and 160 times for the overall IBM calculations compared with the original algorithm while retaining the computational accuracy.
\end{abstract}


\section{Introduction}
The diffuse-interface immersed boundary method (IBM) \citep{peskin2002immersed, lambert2013active} is used in several applications of computational fluid dynamics (CFD) to investigate the generated flow field around geometrically complex and moving objects.
It has been widely used, as it is relatively stable and easy to implement into CFD codes by considering the wall surface represented by a set of Lagrangian points that does not require complicated computational grid generation.
However, one of the inherent bottlenecks for the diffuse-interface IBM has been the lower resolution of the boundary layer, as the external force is diffused around the wall surface.
To overcome this problem, several IBM algorithms have been proposed to enforce the no-slip boundary condition at the wall surface \citep{wu2009implicit, wang2008combined}.
In this regard, a boundary-condition-enforced IBM has been proposed by \citet{pinelli2010immersed}, which imposes a spreading operator correction that implicitly ensures the no-slip boundary condition for the external force of the diffuse-interface IBM.
It is advantageous over other methods regarding the irrelevance of the velocity field around the wall surface to its convergence of the solution.
While this correction succeeds in enforcing the boundary condition at the Lagrangian points of the wall surface, it features a high computational cost for solving the correction operator required for each Lagrangian point.
The correction coefficients are obtained by solving a linear equation iteratively, which scales in the order of $\mathcal{O}(N^2)$ as the number of Lagrangian points, $N$, increases.
In large-scale simulations in which a large number of Lagrangian points is required \citep{saito2024large, posa2021analysis}, the computational cost of solving the correction operator quickly scales up to dozens of times that of the flow field calculation.
Previous studies \citep{valero2015accelerating, boroni2017full} have utilized GPUs to accelerate the calculation, but such an approach cannot be implemented on pure CPU-based computing machines.
Moreover, such an approach lacks scalability for a larger number of Lagrangian points for simulating high-Reynolds-number flow fields, as it has not solved the scalability issue.
This paper presents an algorithm-based approach utilizing a simplified calculation for solving the correction operator, which significantly reduces the computational costs when the wall surface features an extruded geometry while retaining the computational accuracy.

\section{Numrical methods}
\subsection{Boundary-condition-enforced IBM}
When employing the diffuse-interface IBM, the wall surface is represented by a group of Lagrangian points, and using those points, physical quantities on the Eulerian grid are modified by imposing an imaginary external force so that all the physical quantities, such as velocity and pressure, satisfy the boundary conditions at the wall surface. 
The imaginary external force is derived through the following procedure.

First, the equation of momentum conservation is solved without considering the existence of the wall.
Subsequently, the velocity on the Lagrangian points at the wall surface is interpolated from the temporal solution from the Eulerian grid as
\begin{equation}
	\mbox{\boldmath $U_{s}$} = \sum_{j\in D_s}\hat{\mbox{\boldmath $u$}}_j \delta_h \left( \frac{|x_j-X_s|}{\Delta x}\right) \delta_h \left( \frac{|y_j-Y_s|}{\Delta y}\right) \delta_h \left( \frac{|z_j-Z_s|}{\Delta z}\right),
\end{equation}
where $\hat{\mbox{\boldmath $u$}}$ is the temporal solution of velocity before considering the existence of the wall, \mbox{\boldmath $U_{s}$}, ($X_s$, $Y_s$, $Z_s$) are the velocity, density, and the coordinate of the $s^{th}$ Lagrangian point, respectively, and ($x_j, y_j, z_j$), $\Delta x$, $\Delta y$, $\Delta z$ are the $j^{th}$ coordinate of the Eulerian grid and the grid sizes at the coordinates, respectively.
Note that in this analysis, a staggered Cartesian coordinate system is adopted; hence, the coordinates are different among density and velocity.
Furthermore, $D_s$ is the small group of points within the Eulerian grid around the $s^{th}$ Lagrangian point, and $\delta_h (r)$ is represented by
\begin{equation}
  \label{equation:delta}
  \delta_h ( r ) = \left\{
  \begin{array}{ll}
    \frac{1}{3}(1+\sqrt{-3r^2 +1}) & (0 \leq r \leq 0.5)\\
    \frac{1}{6}(5-3r-\sqrt{-3(1-r)^2 +1}) & (0.5 < r \leq 1.5)\\
    0 & otherwise
  \end{array}
  \right.
\end{equation}
Using this interpolated velocity and the velocity of the moving wall at each Lagrangian point, $\mbox{\boldmath $U_{w}$}$, the imaginary external force, $\mbox{\boldmath $F_{s}$}$, at the Lagrangian points is calculated using 
\begin{equation}
  \mbox{\boldmath $F_{s}$} =\rho \frac{\mbox{\boldmath $U_{w}$}-\mbox{\boldmath $U_{s}$}}{\Delta t} .
\end{equation}

\noindent
Once $\mbox{\boldmath $F_{s}$}$ is calculated, it is then interpolated back to the Eulerian grid as
\begin{equation}
	\mbox{\boldmath $f_{s}$} = \sum_{s\in D_j} \mbox{\boldmath $F_{s}$} \delta_h \left( \frac{|x_j-X_s|}{\Delta x}\right) \delta_h \left( \frac{|y_j-Y_s|}{\Delta y}\right) \delta_h \left( \frac{|z_j-Z_s|}{\Delta z}\right)\Delta s_{s} \epsilon_{s},
\end{equation}
where $\Delta s_{s}$ and $\epsilon_{s}$ are the length of the arc joining $\boldmath{X_{s-1/2}}$ and $\boldmath{X_{s+1/2}}$, and the correction operator determined by the positional relationship between the Eulerian grid and the Lagrangian points solving the following equations, respectively.
\begin{equation}
  \label{equation:matrix}
	\mbox{\boldmath $A_{kl}$} = \sum_{j\in D_l} \delta_h \left( \frac{|\mbox{\boldmath $x_{j}$}-\mbox{\boldmath $X_{k}$} |}{\boldmath \Delta}\right) \delta_h \left( \frac{|\mbox{\boldmath $x_{j}$}-\mbox{\boldmath $X_{l}$} |}{\boldmath \Delta}\right)\Delta s_{k}
\end{equation}
\begin{equation}
  \label{equation:epsilon}
	\mbox{\boldmath $A \epsilon$} = 1
\end{equation}
\noindent
The correction operator implicitly ensures the no-sip boundary condition at the wall surface, which was originally proposed by \citet{pinelli2010immersed}.
Finally, the interpolated external force, \mbox{\boldmath $f_{s}$}, is imposed on the Eulerian grid.

\subsection{Matrix simplifications for solving the correction operator, \mbox{\boldmath $\epsilon$}}
\label{sec:headings}

In the original algorithm to solve the correction operator in equations (\ref{equation:matrix}) and (\ref{equation:epsilon}), the computational cost scales in the order of $\mathcal{O}(N^2)$ as the number of Lagrangian points, $N$, representing the wall surface increases.
This is because, when $N$ increases, the size of the matrix $\mbox{\boldmath $A$} \in \mathcal{R}^{NxN}$ also increases, and the computational cost for solving \mbox{\boldmath $\epsilon$} iteratively is directly affected by the size of the matrix \mbox{\boldmath $A$}.  
Therefore, if the size of the matrix \mbox{\boldmath $A$} can be reduced, the computational cost to solve the correction operator can be reduced.

\begin{figure}
  \centering
  \includegraphics[keepaspectratio,width=0.8\linewidth]{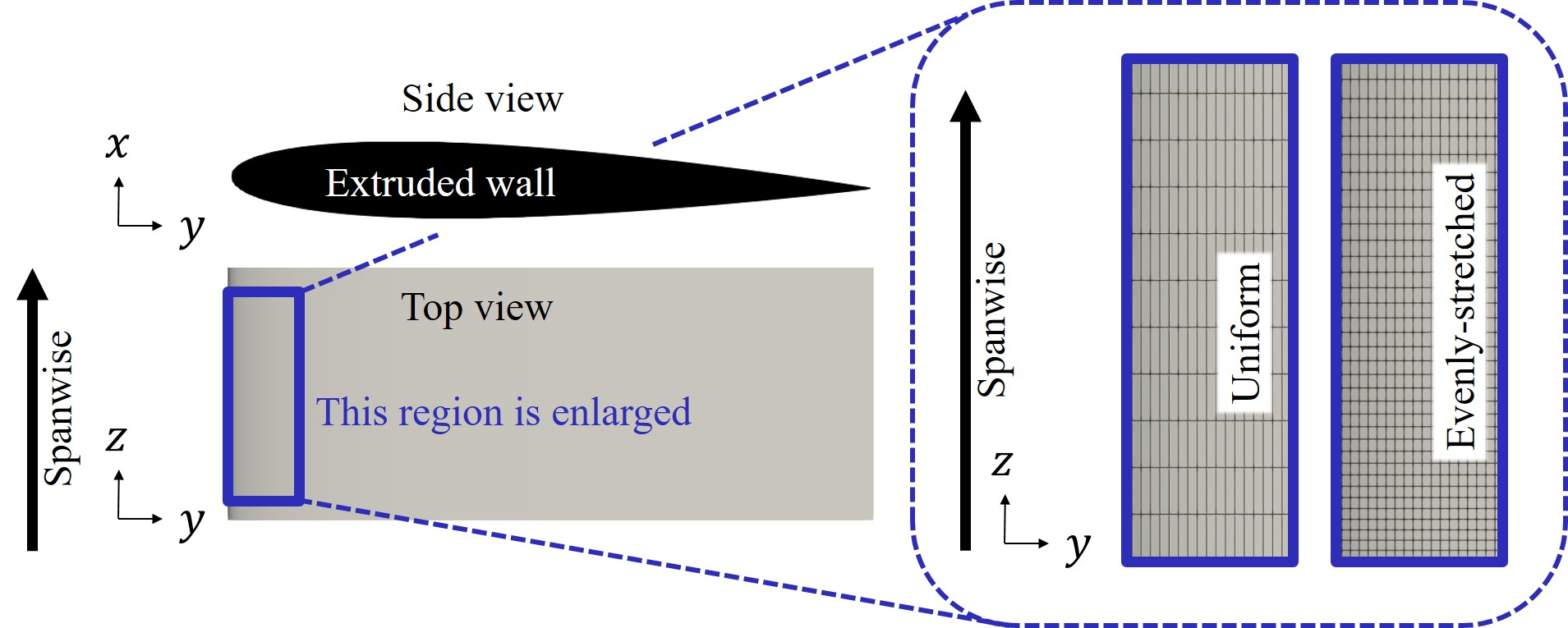}
  \caption{Example of the Lagrangian points configurations that the presented algorithm is applicable (the intersections of the lines represent at which the Lagrangian points are located).}
  \label{fig:lagrangian-euler}
\end{figure}

When the wall surface features a linearly extruded geometry, such as a cylinder or constant-chord wing, the matrix \mbox{\boldmath $A$} can be efficiently simplified to reduce the computational cost.
In the following sections, a constant-chord NACA airfoil is considered as an example.
Figure \ref{fig:lagrangian-euler} shows the distribution of the Lagrangian points on the wall surface along the spanwise direction.
Here, the intersections of the lines represent the locations of the Lagrangian points.
The presented algorithm is applicable to grid points that are uniform or evenly stretched in a spanwise direction, and its applicability to an evenly stretched grid is discussed later. 
In this study, the Lagrangian points are generated and aligned by repeating the same sequence on each consecutive plane perpendicular to the spanwise direction from one end of the span to the other.

\subsubsection{Simplification No.1: Using the symmetricity of the solution of \mbox{\boldmath $\epsilon$}}
\begin{figure}
  \centering
  \includegraphics[keepaspectratio,width=0.5\linewidth]{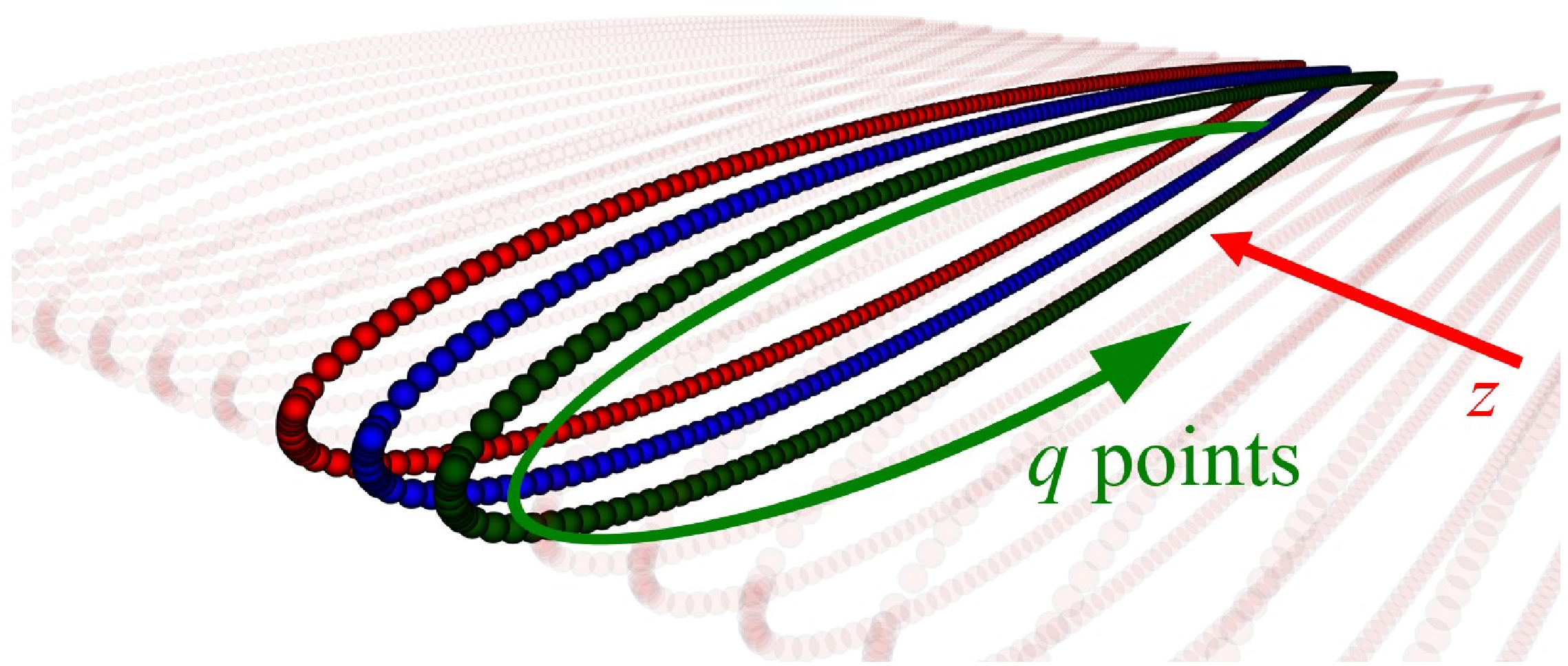}
  \caption{Schematic of the segmentation of the Lagrangian points considering symmetricity of the solution, \mbox{\boldmath $\epsilon$}. Each color represents the Lagrangian points in the same symmetric plane ($q$ denotes the number of Lagrangian points in the same symmetrical plane). }
  \label{fig:simplification1}
\end{figure}

Figure \ref{fig:simplification1} shows the distribution of the Lagrangian points representing the wall surface of the airfoil.
Here, $q$ represents the number of Lagrangian points on one of the $x-y$ planes, and the common color (red, blue, and green) shows that the Lagrangian points are on the same plane.
The Lagrangian points are evenly spaced along the $z$-direction.
By virtue of the patterned relationship between the Lagrangian points and the Eulerian grid, each $A_{kl}$ in equation (\ref{equation:matrix}) features the same value when either the $k^{th}$ and $l^{th}$ Lagrangian points share the same $x$- and $y$-coordinates.
Therefore, focusing on the values of \mbox{\boldmath $\epsilon$} for each Lagrangian point, when the Lagrangian points share the same $x$- and $y$-coordinates, they share the same \mbox{\boldmath $\epsilon$}.
Thus, while the solution of \mbox{\boldmath $\epsilon$} in equations (\ref{equation:matrix}) and (\ref{equation:epsilon}) originally had $N$ degrees of freedom, it is decreased to $N/r$ ($=q$) degrees of freedom by considering the patterned relationship in the coordinates.
Here, $r$ is the number of Eulerian or Lagrangian points in the $z$-direction.
This enables the modification of equation (\ref{equation:epsilon}) to equation (\ref{equation:simplification1}), as the required number of equations can be reduced from $N$ to $q$.
\begin{equation}
  \label{equation:simplification1}
  \centering
  \begin{bmatrix}
  A_{11} & A_{12} & \dots & A_{1N} \\
  A_{21} & \ddots &  & \vdots \\
  \vdots &  & \ddots & \vdots \\
  A_{q1} & A_{N2} & \dots & A_{qN} 
  \end{bmatrix}
  \begin{bmatrix}
  \epsilon_1 \\
  \epsilon_2 \\
  \vdots \\
  \epsilon_N 
  \end{bmatrix}
  =
  \begin{bmatrix}
  1 \\
  1 \\
  \vdots \\
  1 
  \end{bmatrix} 
\end{equation}

\subsubsection{Simplification No.2: Using the sparseness of the matrix \mbox{\boldmath $A$}}
\begin{figure}
  \centering
  \includegraphics[keepaspectratio,width=0.5\linewidth]{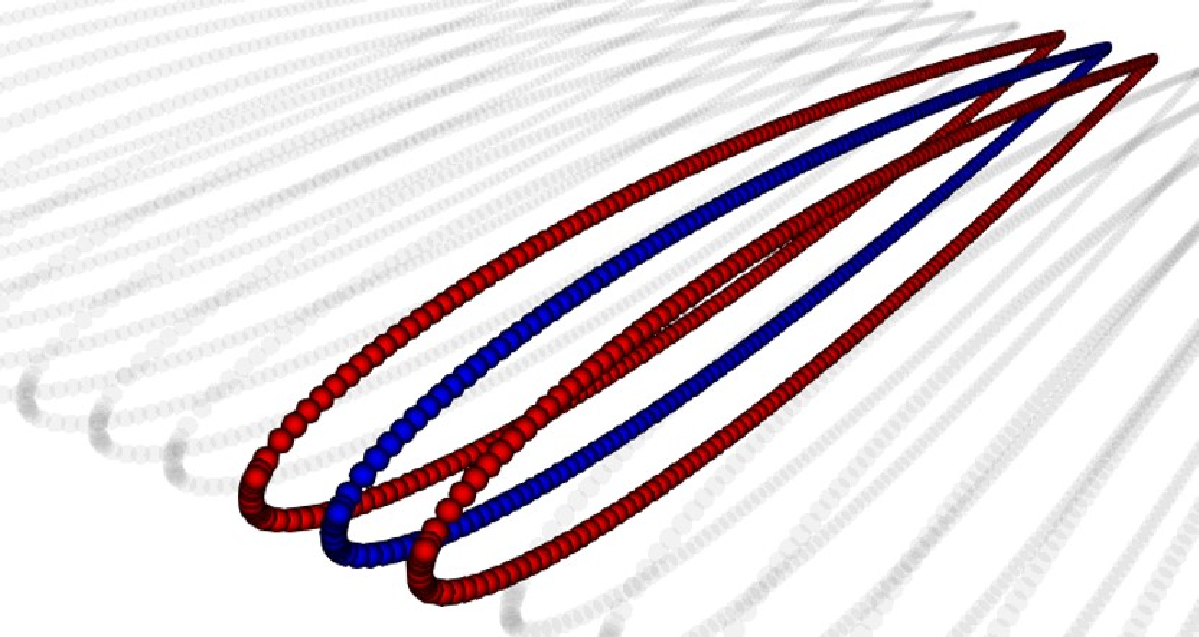}
  \caption{Schematic of the related Lagrangian points when considering the solution of \mbox{\boldmath $\epsilon$} for blue Lagrangian points (blue: interested in its solution; red: consideration required to obtain the solution for blue points; grey: irrelevant).}
  \label{fig:simplification2}
\end{figure}

Figure \ref{fig:simplification2} shows the Lagrangian points related to solving the values of \mbox{\boldmath $\epsilon$} on an arbitral $x-y$ plane.
The blue Lagrangian points are relevant to solving the solution of \mbox{\boldmath $\epsilon$}.
From equation (\ref{equation:delta}), each value of $A_{kl}$ whose $z$-coordinate is more than 1.5 $\Delta z$ away from the red Lagrangian points is null.
From this point of view, in Fig. \ref{fig:simplification2}, only the red Lagrangian points, which are adjacent to the blue Lagrangian points in the $z$-direction, are required to solve the values of \mbox{\boldmath $\epsilon$}.
Therefore, the following equation holds on any $x-y$ plane.
\begin{equation}
  \label{equation:simplification2.1}
  \centering
  \begin{bmatrix}
  A_{11} & A_{12} & \dots & A_{1N} \\
  A_{21} & \ddots &  & \vdots \\
  \vdots &  & \ddots & \vdots \\
  A_{q1} & A_{N2} & \dots & A_{qN} 
  \end{bmatrix}
  \begin{bmatrix}
  \epsilon_1 \\
  \epsilon_2 \\
  \vdots \\
  \epsilon_N 
  \end{bmatrix} 
  =
  \begin{bmatrix}
  \sum_{i=(p-1)q+1}^{(p+2)q} A_{1,i} \epsilon_{i} \\
  \sum_{i=(p-1)q+1}^{(p+2)q} A_{2,i} \epsilon_{i} \\
  \vdots \\
  \sum_{i=(p-1)q+1}^{(p+2)q} A_{N,i} \epsilon_{i} \\
  \end{bmatrix}
\end{equation}
\noindent
Here, $p \in \{1,2,...,r-1\}$, and equation (\ref{equation:simplification2.1}) is satisfied under any $p$.
Then equation (\ref{equation:simplification1}) can be transformed into the following $q \times 3q$ matrix equation.

\begin{equation}
  \label{equation:simplification2.2}
  \centering
  \begin{bmatrix}
  A_{1,(p-1)q+1} & A_{1,(p-1)q+2} & \dots & A_{1,(p+2)q} \\
  A_{2,(p-1)q+1} & \ddots &  & \vdots \\
  \vdots &  & \ddots & \vdots \\
  A_{q,(p-1)q+1} & A_{N2} & \dots & A_{q,(p+2)q} 
  \end{bmatrix}
  \begin{bmatrix}
  \epsilon_{(p-1)q+1} \\
  \epsilon_{(p-1)q+2} \\
  \vdots \\
  \epsilon_{(p+2)q} 
  \end{bmatrix}
  =
  \begin{bmatrix}
  1 \\
  1 \\
  \vdots \\
  1 
  \end{bmatrix} 
\end{equation}

\subsubsection{Simplification No.3: Matrix transformation for Krylov subspace iteration methods}
Krylov subspace iteration methods are capable of solving the original equation (\ref{equation:epsilon}) in a few iterations without any preconditioning \citep{pinelli2010immersed}; hence, the same iteration method is used in this context.
The above simplifications reduce the size of the matrix \mbox{\boldmath $A$} from the original $N \times N$ to $q \times 3q$, but the non-square matrix feature of \mbox{\boldmath $A$} hinders the use of Krylov subspace iteration methods.
To apply Krylov subspace iteration methods, modifying the matrix \mbox{\boldmath $A$} to a square matrix is preferable, and by taking advantage of the following relationship among \mbox{\boldmath $\epsilon$}, the matrix \mbox{\boldmath $A$} can be modified into a $q \times q$ matrix.
\begin{equation}
  \{\epsilon_{(p-1)q+1},\epsilon_{(p-1)q+2},...,\epsilon_{(p+2)q}\}=\{\epsilon_{pq+1},...,\epsilon_{(p+1)q},\epsilon_{pq+1},...,\epsilon_{(p+1)q},\epsilon_{pq+1},...,\epsilon_{(p+1)q}\}
\end{equation}
\begin{equation}
\centering
\begin{bmatrix}
A_{1,(p-1)q+1}+A_{1,pq+1}+A_{1,(p+1)q+1} &  \dots & A_{1,pq}+A_{1,(p+1)q}+A_{1,(p+2)q} \\
\vdots & \ddots & \vdots \\
A_{q,(p-1)q+1}+A_{q,pq+1}+A_{q,(p+1)q+1} &  \dots & A_{q,pq}+A_{q,(p+1)q}+A_{q,(p+2)q}
\end{bmatrix} 
\begin{bmatrix}
\epsilon_{pq+1} \\
\vdots \\
\epsilon_{(p+1)q} 
\end{bmatrix} 
= \begin{bmatrix}
1 \\
\vdots \\
1 
\end{bmatrix}
\end{equation}

The Lagrangian points can also be evenly stretched along the spanwise direction to apply the simplifications, which improves the range of applications.
This is because the matrix \mbox{\boldmath $A$} in equation (\ref{equation:matrix}) is only dependent on the relative distance between the Eulerian grid and the Lagrangian points.
Analytically, such a dependency can be explicitly shown in the following equations.
It is safe to assume that the $z$-coordinates of both the Eulerian grid and the Lagrangian points align, as this is ideally the best and easiest way to locate the Lagrangian points.
The ideal distance of the Lagrangian points being identical to the Eulerian grid size was originally discussed by \citet{pinelli2010immersed}. 
Therefore, the solution of $|z_j-Z_s|$ can be explicitly expressed as one of the following patterns.
\begin{equation}
	|z_j-Z_s| = \left\{
	\begin{array}{ll}
		\Delta z_{j}\\
		0\\
		\Delta z_{j+q} & 
	\end{array}
	\right.
\end{equation}
\noindent
Here, $\Delta z_{j}$ is defined by $|z_{j}-z_{j-q}|$.
As the airfoil consists of $q$ Lagrangian points in each plane perpendicular to the spanwise direction, $\mbox{\boldmath $X_{j}$}$ and $\mbox{\boldmath $X_{j-q}$}$ are adjacent in the $z$-direction and share the same $x$- and $y$-coordinates.
When the grid points are evenly stretched by the stretch rate of $\alpha$ in the spanwise direction, the solution of $|z_j-Z_s|/\Delta z_{j}$ can be written as
\begin{equation}
	\frac{|z_j-Z_s|}{\Delta z_{j}} = \left\{
	\begin{array}{ll}
		1\\
		0\\
		\alpha& 
	\end{array}
	\right.,
\end{equation}
\noindent
as $\Delta z_{j+q}=\alpha \Delta z_{j}$.
This shows the independence of the matrix \mbox{\boldmath $A$} to the local grid size in a spanwise direction, and the simplifications can be conducted in the same manner as the uniform grid case.
Therefore, even when the grid is stretched in a spanwise direction, the solution of equation (\ref{equation:epsilon}) does not change along the spanwise direction.
Note that, in the current algorithm, the stretch rate needs to be relatively moderate for the second adjacent Lagrangian points in the $z$-direction to be out of context.
This limitation is described by satisfying $\alpha^2+\alpha>1.5$ considering that $\delta_h (r)$ in equation (\ref{equation:delta}) considers 1.5$\Delta z$ peripheral Lagrangian points.
Although this results in the maximum stretch rate of 21.5\% in the spanwise direction, it can be considered that this limitation is within the range of practical usage.
For a higher stretch rate, one can transform the matrix \mbox{\boldmath $A$} in a similar way by considering additional Lagrangian points in the spanwise direction.
For reference, an additional consideration of Lagrangian points on the grid's shrinking side in the spanwise direction can increase the limitation of the stretch rate to 44.6\%, which is derived from $\alpha^3+\alpha^2+\alpha>1.5$.

\section{Results and discussion}
\subsection{Computational efficiency and accuracy analyses}
\begin{figure}
	\captionsetup[subfloat]{labelformat=simple}
	\centering
	\subfloat[Computational domain (Dotted arrow denotes plunging movement of the airfoil).\label{fig:domain1}]{{\includegraphics[width=0.45\linewidth]{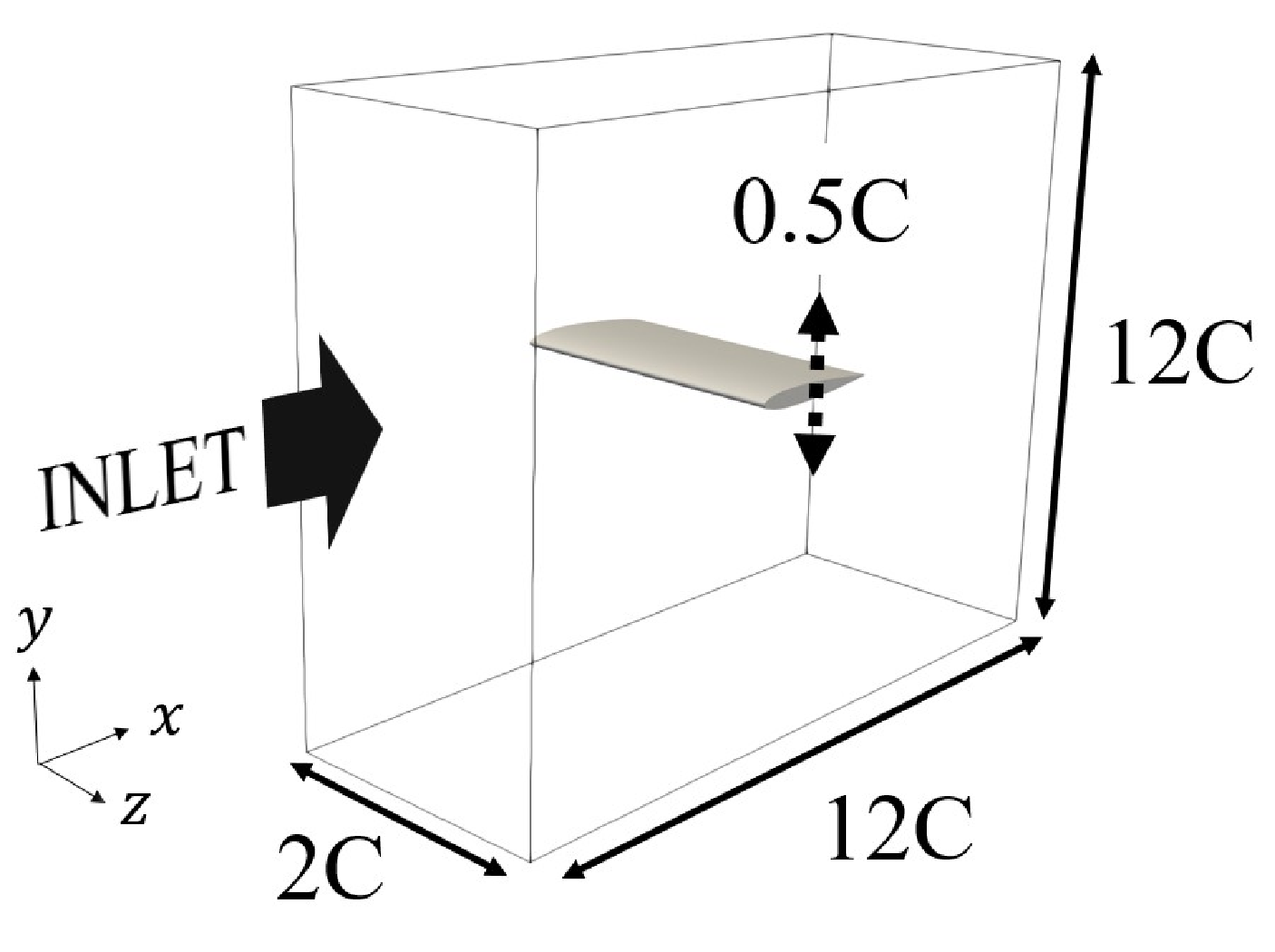} }}
	\subfloat[Nondimensionalized vorticity contour\label{fig:2D-flow}]{{\includegraphics[width=0.4\linewidth]{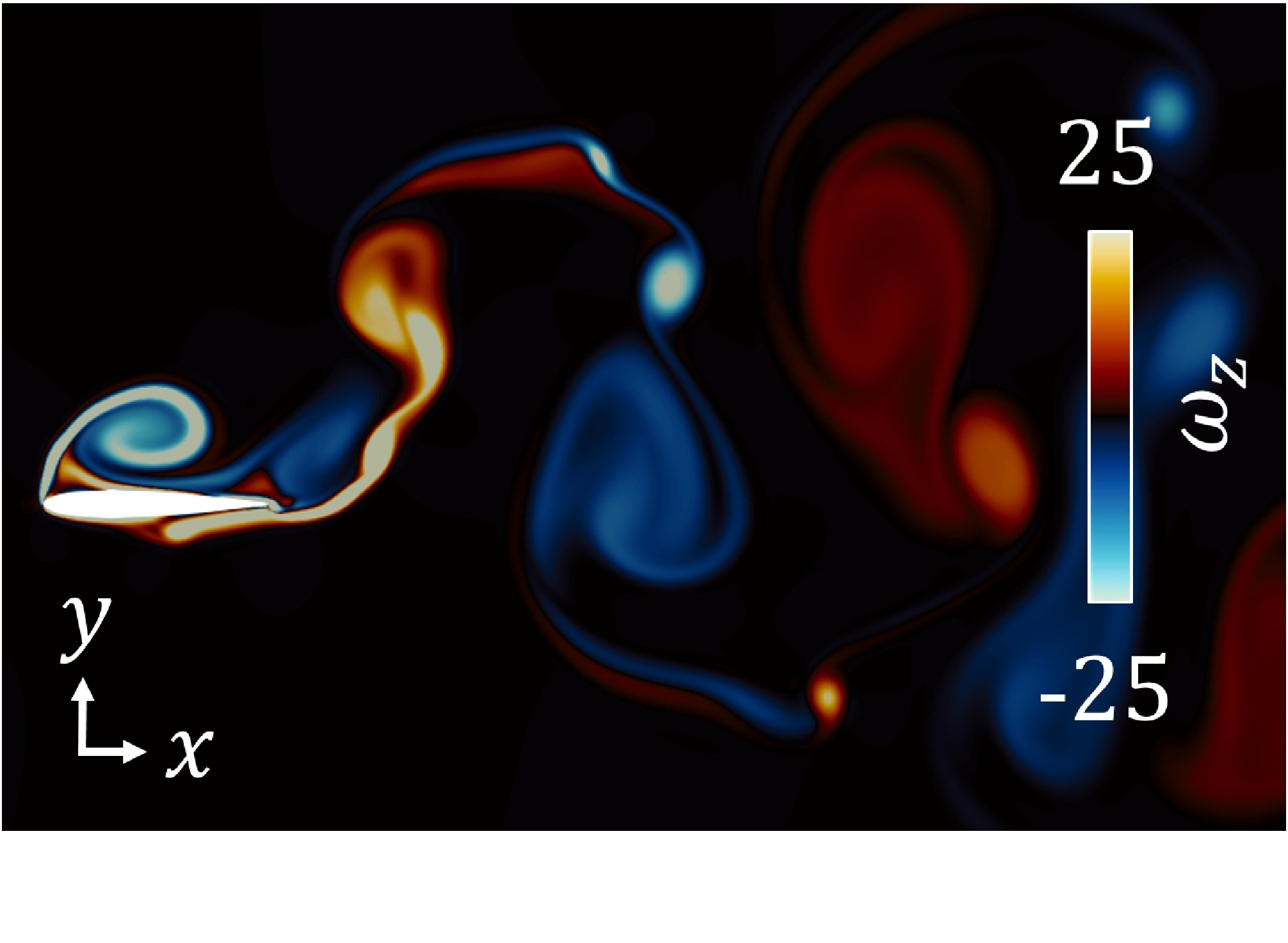} }}
	\caption{Computational domain and the result of the numerical simulation of a plunging airfoil.}
	\label{fig:LES}
\end{figure}

To evaluate the improvement in the computational efficiency of the boundary-condition-enforced IBM, large-eddy simulations (LES) of a plunging airfoil were conducted by solving the incompressible Navier-Stokes equations.
The computational domain and calculation result are shown in Fig. \ref{fig:LES}.
The airfoil features the NACA0010 configuration, and based on the chord length of the airfoil, C, the size of the computational domain is 12C, 12C, and 2C in the $x$-, $y$-, and $z$- directions and the number of grid points is 1140, 550, and 67, respectively.
The spanwise length in the $z$-direction is 2C, and the boundary conditions are outflow conditions in the $x$- and $y$- directions, respectively, and periodic in the $z$-direction.
The minimum grid size is ($\Delta x, \Delta y, \Delta z$) = (0.006C, 0.006C, 0.03C).
The inlet flow is imposed at the negative end of the $x$-direction.
The Reynolds number is based on the ambient flow velocity, $U$, and the chord length of the airfoil is 3,000.
The plunging airfoil features the sinusoidal oscillation in the $y$-direction with a plunging amplitude of 0.5C and a frequency of $0.5U$/C Hz with a zero-degree angle of attack relative to the ambient flow.
The airfoil consists of $332(=q)$ Lagrangian points on each $x-y$ plane with a total of $22,244(=N)$ points in the entire domain.
The numerical simulation is performed using an in-house solver referred to as FK$^3$ (e.g. \cite{pillai2022numerical, kai2023flamelet}), which consists of a fractional-step method that employs a pressure-based semi-implicit algorithm \citep{moureau2007efficient}.
The unresolved subgrid-scale modeling was conducted using the dynamic Smagorinsky model \citep{lilly1992proposed}.
In Fig. \ref{fig:2D-flow}, the vorticity is normalized by the plunging frequency of the airfoil.

\begin{figure}
	\captionsetup[subfloat]{labelformat=simple}
	\centering
	\subfloat[Accuracy\label{fig:epsilon}]{{\includegraphics[width=0.45\linewidth]{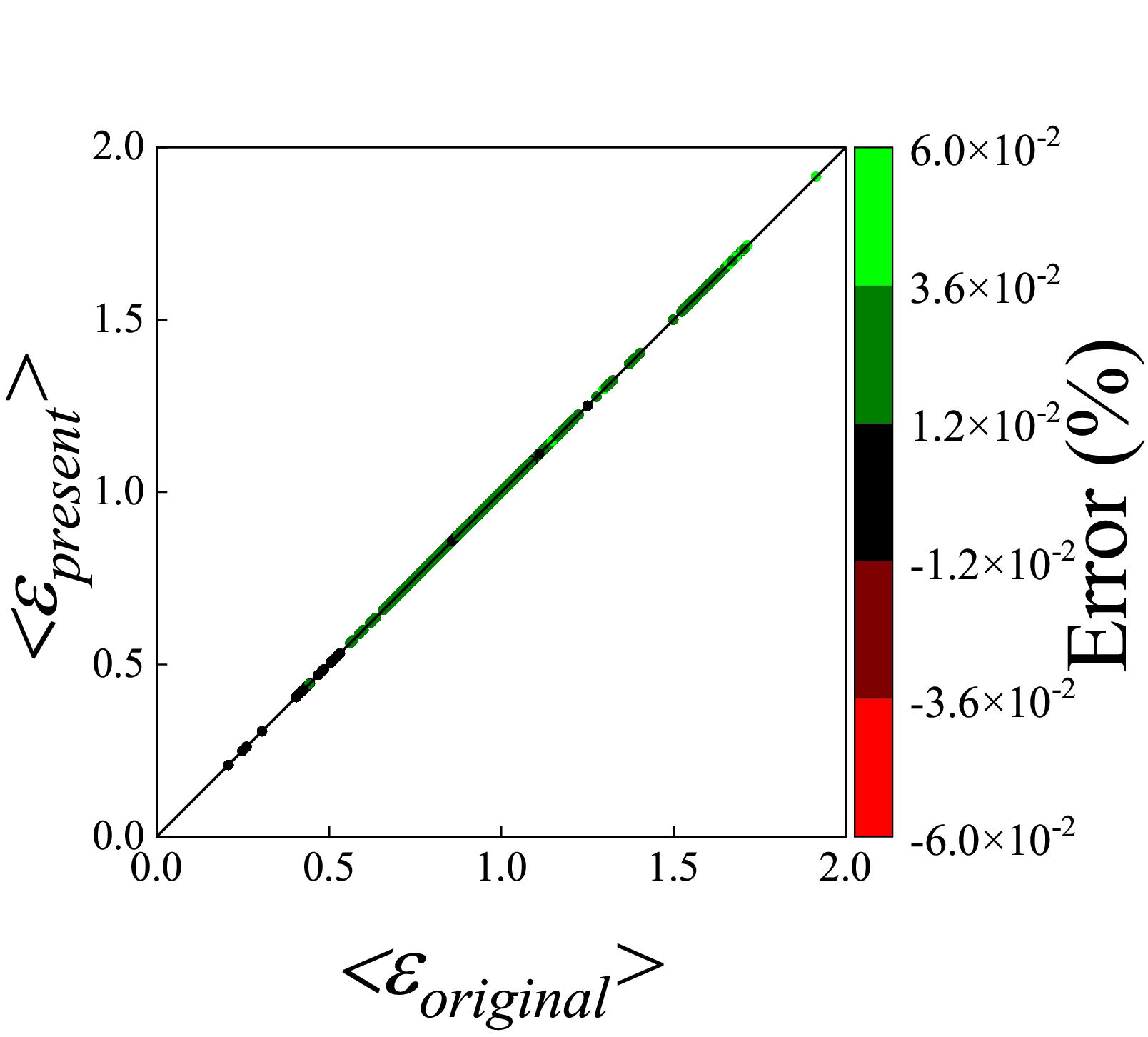} }}
	\subfloat[Computational cost (log scale)\label{fig:computational_cost}]{{\includegraphics[width=0.5\linewidth]{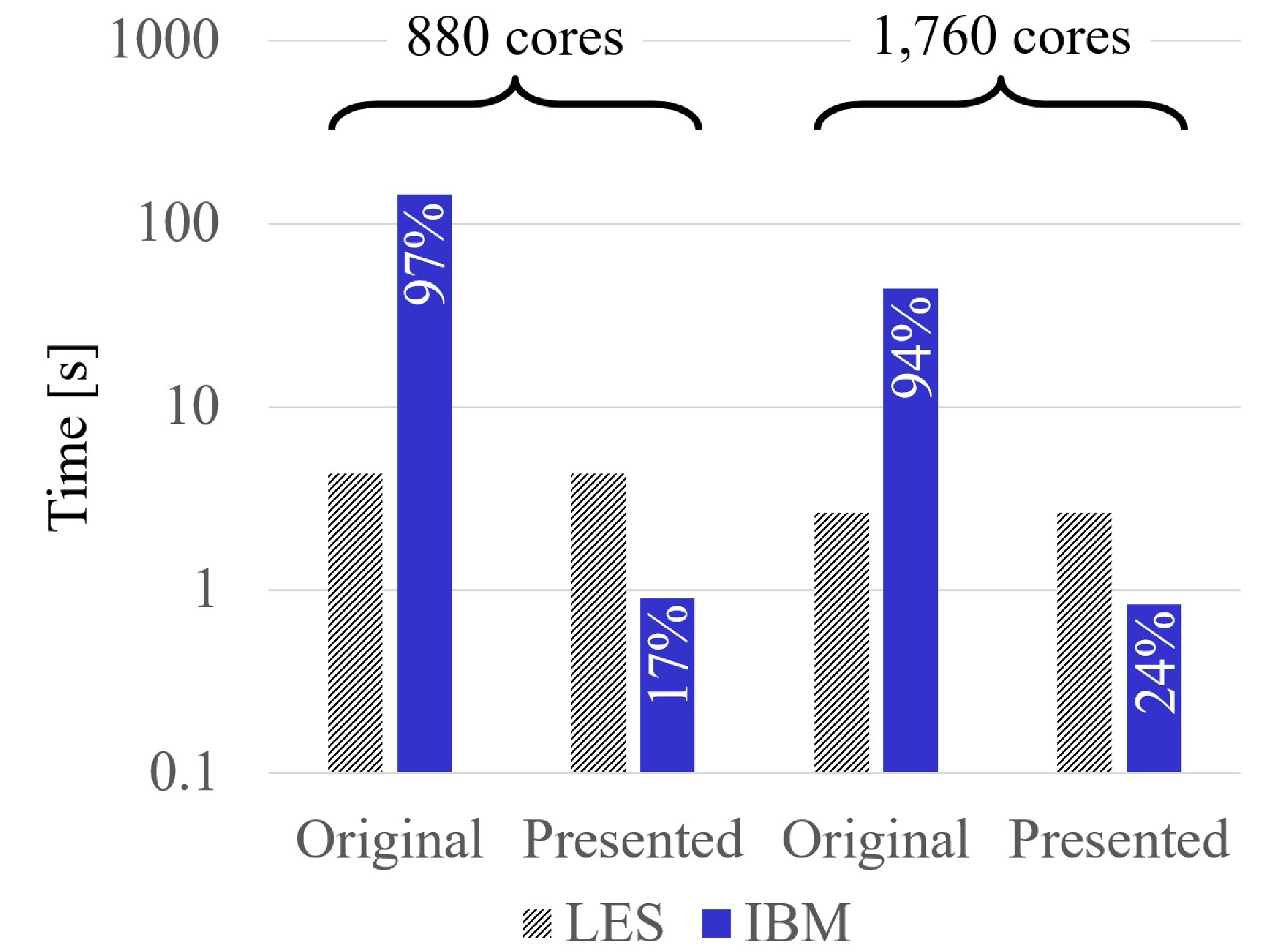} }}
	\caption{Comparisons of (a) the solution of the normalized correction operator, \mbox{\boldmath $\epsilon$} (the color scale denotes ($\epsilon_{present}-\epsilon_{original})/\overline{\epsilon_{original}}\times100$), and (b) the computational time between original and presented IBM algorithms (the percentages indicate the proportion of the computational cost of IBM).}
	\label{fig:comparisons}
\end{figure}

Figure \ref{fig:comparisons} shows the comparisons of the accuracy and the computational efficiency between the original and presented algorithms.
Figure \ref{fig:epsilon} shows the comparison of the correction operator, \mbox{\boldmath $\epsilon$}, normalized by the averaged value of \mbox{\boldmath $\epsilon$} from the original IBM algorithm between the original and presented IBM algorithms.
Approximately 10,000 points were sampled from various locations in the airfoils, and time steps were plotted.
The colored scale indicates the error value obtained from ($\epsilon_{present}-\epsilon_{original})/\overline{\epsilon_{original}}\times100$ for each Lagrangian point, where $\overline{\epsilon_{original}}$ is the averaged value of $\epsilon_{original}$, and the magnitude of the largest discrepancy is 0.058\%, which is significantly small.
Here, the iterative convergence strategy of the bi-conjugate gradient stabilized (BiCGSTAB) method \citep{van1992bi} was used.

Subsequently, the computational cost was compared between the presented and the original IBM algorithms.
The calculation was parallelized using 880 and 1,760 CPU cores for the comparison.
Figure \ref{fig:computational_cost} shows the computational cost for each step and its breakdown among the IBM and LES on a logarithmic scale.
While the computational costs of the IBM employing the previous algorithm accounted for 94\% to 97\% of the overall cost, those of the IBM employing the presented algorithm only accounted for 17\% to 24\% of the overall cost, resulting in 52 to 160 times the speed-up of the IBM calculation.
In fact, the computational speed-up predominantly resulted from the above-described simplifications of the correction operator calculation, which alone resulted in a speed-up of 800 to 2,800 times. 
Therefore, the evaluation of the correction operator, \mbox{\boldmath $\epsilon$}, and the computational speed-ups indicate that the presented algorithm is capable of reproducing the solution of \mbox{\boldmath $\epsilon$} in a significantly efficient and accurate manner.

\subsection{Scalability analysis}
To assess the scalability of the presented algorithm, its computational complexity is compared with that of the previous algorithms in Table \ref{tab:complexity}.
Here, $r$ and $\kappa$ represent the number of Eulerian or Lagrangian points in the $z$-direction and the condition number, respectively.
The complexity of $\mathcal{O}(max(N, N^2/r^2))$ in the boundary-condition-enforced IBM for an extruded wall geometry is derived from the fact that the original correction operator calculation, which had the computational complexity of $\mathcal{O}(N^{2})$, is simply reduced by $r \times r$ owing to the size reduction of the matrix $A$ from $\mbox{\boldmath $A$} \in \mathcal{R}^{NxN}$ to $\mbox{\boldmath $A_{reduced}$} \in \mathcal{R}^{qxq}$ ($q=N/r$).
However, the underlying computational complexity of $\mathcal{O}(N)$ cannot be avoided, as this complexity originates from the calculation for each Lagrangian point, such as interpolation from the Eulerian grid to Lagrangian points or the imposition of an external imaginary force.

\begin{table}
  \caption{Comparison of the computational complexity among boundary-codition-enforced IBM algorithms.}
  \begin{center}
  \def~{\hphantom{0}}
  \begin{tabular}{ll}
    \toprule
       Explicit technique-based IBM \citep{zhao2021efficient}                & $\mathcal{O}(N)$                                                               \\
       Multi-direct forcing IBM \citep{luo2007full}                          & $\mathcal{O}(N \sqrt{\kappa})$                                                 \\
       Conjugate gradient technique-based IBM \citep{zhao2021efficient}      & $\mathcal{O}(N \kappa)$                                                        \\
       Boundary condition-enforced IBM \citep{pinelli2010immersed}           & $\mathcal{O}(N^{2})$                                                           \\
       Boundary condition-enforced IBM for extruded wall geometry (presented)   & $\mathcal{O}(max(N, N^2/r^2))$                                                 \\
    \bottomrule
  \end{tabular}
  \label{tab:complexity}
  \end{center}
\end{table}

\begin{figure}
	\captionsetup[subfloat]{labelformat=simple}
	\centering
	\subfloat[Homogeneous scaling\label{fig:computational_time1}]{{\includegraphics[width=0.5\linewidth]{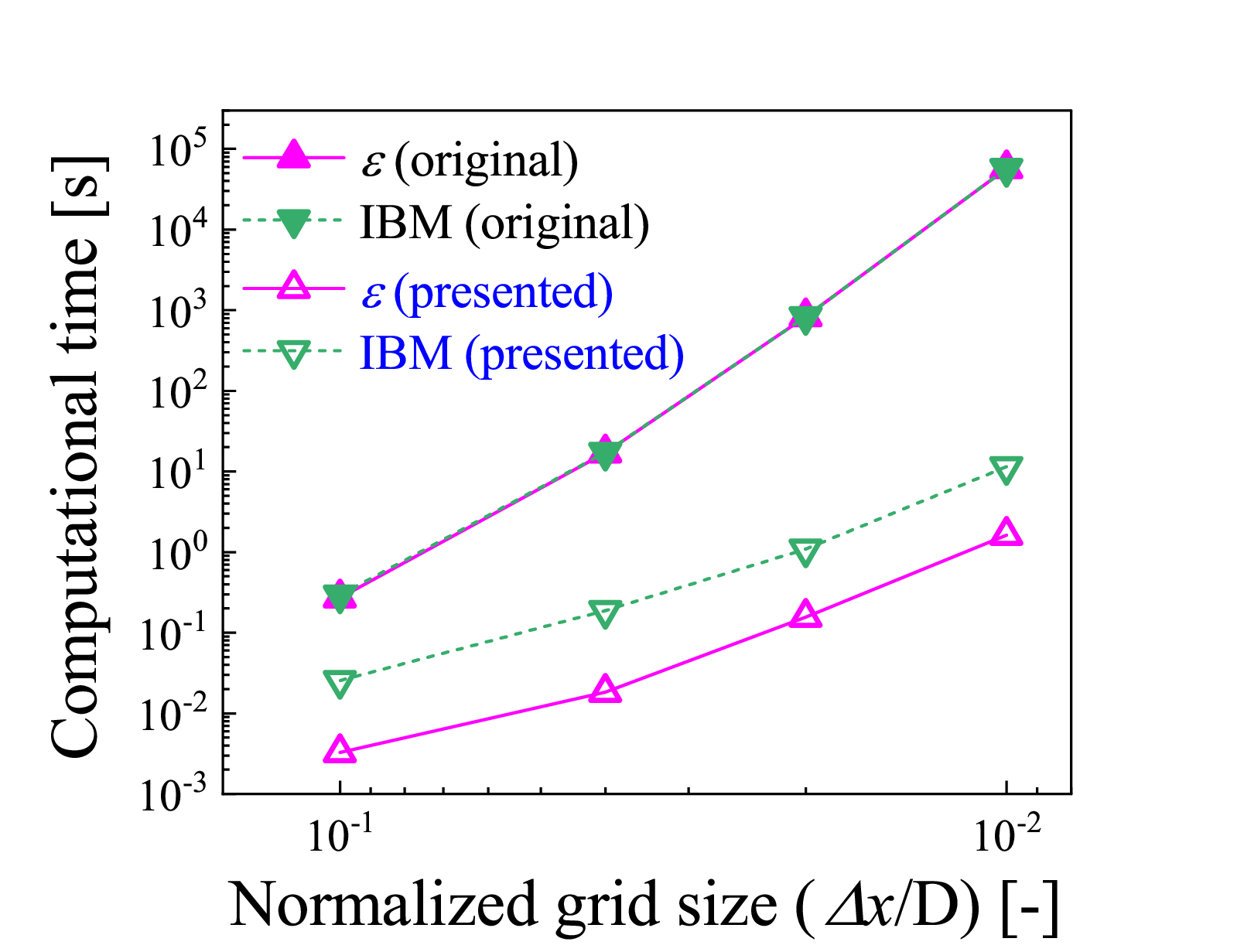} }}
	\subfloat[Spanwise scaling\label{fig:computational_time2}]{{\includegraphics[width=0.5\linewidth]{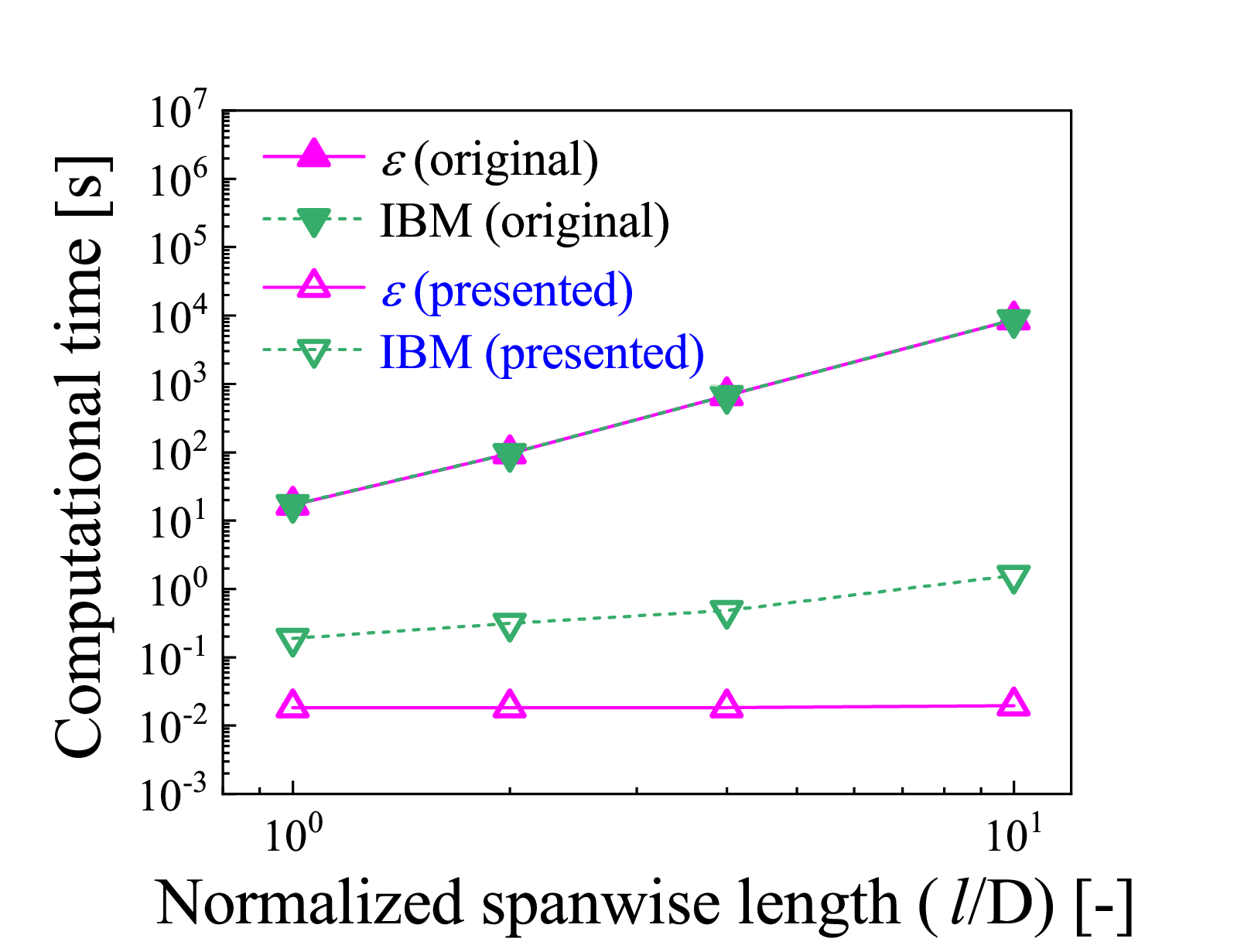} }}
	\caption{Comparison of the computational cost of IBM under (a) homogeneously increased resolution or (b) increased spanwise length, $l$ (pink line and green dashes represent the computational cost of solving the correction operator alone and IBM, respectively, and filled and hollow symbols represent original and presented algorithms, respectively).}
	\label{fig:computational_time}
\end{figure}

The quantitative scalability of the presented algorithm was investigated through attempts to simulate the flow around a circular cylinder under various grid resolutions and spanwise lengths using a single processor without parallelization.
In the original algorithm presented by \citet{pinelli2010immersed}, the resolutions of the Eulerian and Lagrangian points were ideally maintained the same.
Therefore, in this analysis, as the number of Eulerian points increases, so does the number of Lagrangian points accordingly.
Figure \ref{fig:computational_time} shows the comparisons of the computational cost between the original and presented IBMs for different (a) resolutions (homogeneous scaling) and (b) spanwise lengths, $l$ (spanwise scaling).
Homogeneous scaling increases the resolution of the Eulerian grid and Lagrangian points at the same rate in the $x-$, $y-$, and $z-$ directions.
The computational cost of the correction operator calculation alone is also explicitly shown.
Here, the spanwise length and Eulerian grid size are normalized by the diameter of the circular cylinder, D.
In Fig. \ref{fig:computational_time1}, when the resolution of the Eulerian grid and Lagrangian points increases, the computational cost for the correction operator is consistently smaller in the presented algorithm than in the original algorithm by $10^3$ to $10^4$ times.
Moreover, while the correction operator calculation dominates the computational cost in the original algorithm, most of the computational cost in the presented algorithm originates from the calculation other than the correction operator calculation, such as imposing an external forcing term to the velocity field from each Lagrangian point.
This indicates that the use of the presented algorithm almost eliminates the computational cost of the correction operator calculation, which was originally the drawback in the boundary-condition-enforced IBM.
Note that, in practical applications, the parallelization of the code can also be applied to the algorithm to decrease the computational cost further by distributing the Eulerian grid and Lagrangian points among multiple processors, as in the original algorithm.

Furthermore, the presented algorithm is more effective when the number of grid points in a spanwise direction increases by virtue of the geometrical symmetricity.
Figure \ref{fig:computational_time2} shows the comparison of the computational cost of the IBM among different spanwise lengths with a constant grid resolution (i.e. different numbers of grid points in the spanwise direction).
Notably, this feature can be altered to increase the resolution in the spanwise direction with a constant spanwise length.
The computational cost of the original algorithm increases as the spanwise length increases owing to the increase in the Eulerian grid and Lagrangian points.
On the other hand, the computational cost of the presented algorithm does not scale as much as that of the original one.
In fact, the computational cost of the correction operator calculation almost remains constant.

\begin{table}
  \caption{Comparison of the computational cost scaling among IBM algorithms.}
  \begin{center}
  \def~{\hphantom{0}}
  \begin{tabular}{ll}
    \toprule
       Explicit technique-based IBM \citep{zhao2021efficient}                & $\mathcal{O}(N^{1.014})$                                                       \\
       Multi-direct forcing IBM \citep{zhao2021efficient}                    & $\mathcal{O}(N^{1.683})$                                                       \\
       Conjugate gradient technique-based IBM \citep{zhao2021efficient}      & $\mathcal{O}(N^{1.428})$                                                       \\
       Boundary condition-enforced IBM \citep{zhao2021efficient}             & $\mathcal{O}(N^{2.012})$                                                       \\
       Boundary condition-enforced IBM for extruded wall geometry (presented)   &                                                                                \\
       \hspace{5em} (a) homogeneous scaling                                   & $\mathcal{O}(N^{1.352})$                                                       \\
       \hspace{5em} (b) spanwise scaling                                      & $\mathcal{O}(N^{1.008})$                                                       \\
       \bottomrule
  \end{tabular}
  \label{tab:scale}
  \end{center}
\end{table}

The quantitative comparison of the scalability among different IBM algorithms is shown in Fig. \ref{tab:scale}.
The presented algorithm can be evaluated under (a) homogeneous scaling and (b) spanwise scaling.
Spanwise scaling only increases the Eulerian grid and Lagrangian points in a spanwise direction, increasing either the resolution or the domain size.
In both scalings of the presented algorithm, it outperforms all the previously presented algorithms in terms of scalability except for the explicit-technique-based algorithm, which gains computational efficiency by compromising the accuracy.
Furthermore, (b) the spanwise scaling of the presented algorithm is almost identical to that of the explicit-technique-based algorithm.

\section{Conclusions}
A fast algorithm-based calculation algorithm was presented for the boundary-condition-enforced IBM utilizing a simplified calculation for solving the correction operator that significantly reduces the computational costs when the wall surface features an extruded geometry while retaining the computational accuracy.
The algorithm focused on the symmetricity of the wall geometry to simplify the time-consuming matrix calculation, and its computational efficiency and accuracy were evaluated.
The results showed improvements in computational efficiency of up to 160 times for the overall IBM calculations with an error no greater than 0.058\% in the solution.
Furthermore, the scaling of the computational cost consistently remained lower than that of the previously proposed algorithms that do not compromise the computational accuracy.



\section*{Declaration of interests}
The authors report no conflict of interest.





\bibliographystyle{jfm}

\end{document}